\documentclass[twocolumn,secnumarabic,nobibnotes, aps, prb, superscriptaddress,floatfix]{revtex4-2}
\usepackage[normalem]{ulem} 	%need for strikethrough
\usepackage[usenames,dvipsnames]{color}
\usepackage{upgreek}
\usepackage{braket}
\usepackage{float}
\usepackage[nointegrals]{wasysym}
\usepackage{helvet}
\usepackage{hyperref}
\usepackage{xcolor}
\usepackage[english]{babel}
\usepackage{graphicx, import}
\usepackage[export]{adjustbox}
\usepackage{amssymb,amsfonts,amsmath} 
\usepackage{physics}
\usepackage{units}

\usepackage{tikz}
\usetikzlibrary{positioning}

\definecolor{pythonBlue}{rgb}{0.12156862745098039, 0.4666666666666667, 0.7058823529411765}
\definecolor{pythonOrange}{rgb}{1.0, 0.4980392156862745, 0.054901960784313725}

\begin{document}

%////////////////
\title{Tunable Anderson Localization of Dark States}
\author{Jan David Brehm}
	\affiliation{Physikalisches Institut, Karlsruhe Institute of Technology, 76131 Karlsruhe, Germany}
\author{Paul Pöpperl}
	\affiliation{Institut für Theorie der Kondensierten Materie, Karlsruhe Institute of Technology, 76128 Karlsruhe, Germany}
	\affiliation{Institute for Quantum Materials and Technologies, Karlsruhe Institute of Technology, 76021 Karlsruhe, Germany}
\author{Alexander D. Mirlin}
	\affiliation{Institut für Theorie der Kondensierten Materie, Karlsruhe Institute of Technology, 76128 Karlsruhe, Germany}
	\affiliation{Institute for Quantum Materials and Technologies, Karlsruhe Institute of Technology, 76021 Karlsruhe, Germany}
	\affiliation{Landau Institute for Theoretical Physics, 119334 Moscow, Russia}
\author{Alexander Shnirman}	
    \affiliation{Institut für Theorie der Kondensierten Materie, Karlsruhe Institute of Technology, 76128 Karlsruhe, Germany}
    \affiliation{Institute for Quantum Materials and Technologies, Karlsruhe Institute of Technology, 76021 Karlsruhe, Germany}
\author{Alexander Stehli}
	\affiliation{Physikalisches Institut, Karlsruhe Institute of Technology, 76131 Karlsruhe, Germany}
\author{Hannes Rotzinger}
	\affiliation{Physikalisches Institut, Karlsruhe Institute of Technology, 76131 Karlsruhe, Germany}
		\affiliation{Institute for Quantum Materials and Technologies, Karlsruhe Institute of Technology, 76021 Karlsruhe, Germany}
\author{Alexey V. Ustinov}
	\affiliation{Physikalisches Institut, Karlsruhe Institute of Technology, 76131 Karlsruhe, Germany}
		\affiliation{Institute for Quantum Materials and Technologies, Karlsruhe Institute of Technology, 76021 Karlsruhe, Germany}
	\affiliation{National University of Science and Technology MISIS, Moscow 119049, Russia}
	\affiliation{Russian Quantum Center, Skolkovo, Moscow 143025, Russia}

\date{\today}

\begin{abstract}
Random scattering of photons in disordered one-dimensional solids gives rise to an exponential suppression of transmission, which is known as Anderson localization. Here, we experimentally study Anderson localization in a superconducting waveguide quantum electrodynamics system comprising eight individually tunable qubits coupled to a photonic continuum of a waveguide. Employing the qubit frequency control, we artificially introduce frequency disorder to the system and observe an exponential suppression of the transmission coefficient in the vicinity of its subradiant dark modes. The localization length decreases with the disorder strength, which we control in-situ by varying individual qubit frequencies. Employing a one-dimensional non-interacting model of coupled qubits and photons, we are able to support and complement the experimental results. The difference between our investigation and previous studies of localization in qubit arrays is the coupling via a common waveguide, allowing us to explore the localization of mediating photons in an intrinsically open system. The experiment opens the door to the study of various localization phenomena on a new platform, which offers a high degree of control and readout possibilities.

\end{abstract}

\maketitle

\section{Introduction}
The exponential suppression of diffusive transport due to disorder-induced random scattering is known as Anderson localization \cite{Anderson1958,50_years_of_localization}.
Whereas the phenomenon was first considered in electronic systems, where the electronic wave function gets localized due to quantum interference, it proved to be present in a much wider range of different physical systems. This includes localization of light in the optical~\cite{Wiersma1997,Storzer2006} and microwave domain~\cite{Chabanov2000} as well as of sound waves~\cite{Graham1990}.

An emerging platform to study the interactions between photons and matter is superconducting waveguide quantum electrodynamics (wQED), where qubits interact with a broad continuum of electromagnetic modes (rather than with a single mode as in cavity QED) \cite{Gu2017}. Pioneering experiments with single qubits demonstrated, that such systems can be well controlled, feature strong coupling~\cite{astafiev_resonance_2010} and various quantum optical effects like anti-bunched light~\cite{hoi_microwave_2013} and the Autler-Townes splitting~\cite{abdumalikov_electromagnetically_2010, hoi_microwave_2013}. If multiple qubits are coupled to the same waveguide, they obtain an effective photon-mediated infinite-range interaction ~\cite{lalumiere_input-output_2013, loo_photon-mediated_2013}. This gives rise to collective super- and subradiant polaritonic excitations, which correspond to the eigenmodes of finite wQED systems~\cite{zhang_theory_2018,zhong_photon-mediated_2020,brehm_waveguide_2020,loo_photon-mediated_2013}. First experiments showcased a scaled superconducting wQED system featuring multiple qubits in a metamaterial configuration~\cite{brehm_waveguide_2020}, an atomic mirror~\cite{mirhosseini_cavity_2019} and a topological waveguide~\cite{Kim2021}.

Recently it was shown that Anderson localization of light is expected in wQED systems with either position or frequency disorder of the constituent qubits \cite{haakh_polaritonic_2016,Mirhosseini2018,Mirza2018,witthaut_photon_2010,fayard2021manybody}. Although medium-sized wQED systems with controllable qubits can be tuned to a barely disordered state \cite{brehm_waveguide_2020}, they can also be used as a simulator for tunable disorder, which is unavoidably introduced in large-scale superconducting wQED systems lacking local qubit control. This makes them a particularly well suited system to study Anderson localization.

Several recent experimental studies focused on many-body localization in a cavity QED architecture with either directly nearest neighbour coupled qubits  \cite{martinis_mbl_1, martinis_mbl_2,exp_mbl_chen} or indirectly coupled qubits via a bus resonator \cite{exp_mbl_Guo_2020,  exp_mbl_xu}. While these systems are intrinsically closed, here, we investigate localization in an open quantum system and measure propagating photons rather than excitations in the stationary qubits. This provides the benefit to study localization phenomena in a dissipative environment.
Having photons as mediators with a continuous spectrum and directly measuring the transport through the system sets our work apart from previous studies. 

We experimentally investigate Anderson localization in a superconducting wQED system in the non-interacting (low-power) limit. Our system is formed by eight transmon qubits coupled to the mode continuum of a coplanar waveguide. First, we theoretically motivate that such a system can indeed feature Anderson localization in a disordered state by investigating a model of the experiment, which reproduces its main features. Our analysis shows that in a finite system the localization manifests as the suppression of subradiant modes. Based on these insights, we establish the localization experimentally, using the qubits' tunability to introduce frequency disorder to the system. The effective localization length is accessed by performing an average over several random disorder realizations. The experiment features the benefit of precise and in-situ control of the disorder strength and is thus a useful tool for the study of Anderson localization in photonic systems. Due to the non-linearity of the employed superconducting qubits our experiment paves way towards more evolved concepts such as the study of many-body localization phenomena~\cite{mbl_gmp,mbl_baa,mbl_bloch,mbl_rev_nh,mbl_rev_av,mbl_rev_ap}, which were recently discussed in a similar context in Ref.~\cite{fayard2021manybody}.

\section{Theoretical model}
\label{sec:theory}

    \begin{figure}[tb]
        \begin{tikzpicture}[mynode/.style={font=\color{white}\sffamily,fill=pythonBlue,circle,inner sep=1pt,minimum size=0.5cm}, mynode2/.style={font=\color{black}\sffamily,fill=pythonOrange,inner sep=1pt,minimum size=0.5cm}]
            \node[mynode] (a) at (0, 0) {\(0\)};
            \node[mynode, right = 0.7cm of a] (b) {\(0\)};
            \node[mynode, right = 0.7cm of b] (c) {\(0\)};
            \node[mynode, right = 0.7cm of c] (d) {\(0\)};
            \node[mynode, right = 0.7cm of d] (e) {\(0\)};
            \node[mynode, right = 0.7cm of e] (f) {\(0\)};
            \node[mynode, right = 0.7cm of f] (g) {\(0\)};
            % \node[mynode, right = 0.5cm of g] (h) {\(0\)};
            % \node[mynode, right = 0.5cm of h] (i) {\(0\)};
            % \node[mynode, right = 0.5cm of i] (j) {\(0\)};
            % \node[mynode, right = 0.5cm of j] (k) {\(0\)};
            % \node[mynode, right = 0.5cm of k] (l) {\(0\)};
            % \node[mynode, right = 0.5cm of l] (m) {\(0\)};
            % \node[mynode, right = 0.5cm of m] (n) {\(0\)};
            % \node[mynode, right = 0.5cm of n] (o) {\(0\)};
            \node[mynode2, above = 0.7cm of a] (ap) {\(\omega_0\)};
            \node[mynode2, above = 0.7cm of c] (cp) {\(\omega_1\)};
            \node[mynode2, above = 0.7cm of e] (ep) {\(\omega_2\)};
            \node[mynode2, above = 0.7cm of g] (gp) {\(\omega_3\)};
            % \node[mynode2, above = 0.5cm of i] (ip) {\(\omega_4\)};
            % \node[mynode2, above = 0.5cm of k] (kp) {\(\omega_5\)};
            % \node[mynode2, above = 0.5cm of m] (mp) {\(\omega_6\)};
            % \node[mynode2, above = 0.5cm of o] (op) {\(\omega_7\)};
            \draw (a) -- (b) node [midway, above] {\(J\)};
            \draw (b) -- (c) node [midway, above] {\(J\)};
            \draw (c) -- (d) node [midway, above] {\(J\)};
            \draw (d) -- (e) node [midway, above] {\(J\)};
            \draw (e) -- (f) node [midway, above] {\(J\)};
            \draw (f) -- (g) node [midway, above] {\(J\)};
            % \draw (g) -- (h) node [midway, above] {\(J\)};
            % \draw (h) -- (i) node [midway, above] {\(J\)};
            % \draw (i) -- (j) node [midway, above] {\(J\)};
            % \draw (j) -- (k) node [midway, above] {\(J\)};
            % \draw (k) -- (l) node [midway, above] {\(J\)};
            % \draw (l) -- (m) node [midway, above] {\(J\)};
            % \draw (m) -- (n) node [midway, above] {\(J\)};
            % \draw (n) -- (o) node [midway, above] {\(J\)};
            \draw (a) -- (ap) node [midway, right] {\(g\)};
            \draw (c) -- (cp) node [midway, right] {\(g\)};
            \draw (e) -- (ep) node [midway, right] {\(g\)};
            \draw (g) -- (gp) node [midway, right] {\(g\)};
            % \draw (i) -- (ip) node [midway, right] {\(g\)};
            % \draw (k) -- (kp) node [midway, right] {\(g\)};
            % \draw (m) -- (mp) node [midway, right] {\(g\)};
            % \draw (o) -- (op) node [midway, right] {\(g\)};
        \end{tikzpicture}
        \caption{Visualization of the Hamiltonian~\eqref{eq:full_hamilton_site_space} with \(N=4\) qubits, \(N_{\rm int}=1\), and  \(N_\gamma = (N_{\rm int} + 1) N - N_{\rm int}=7\) photon sites. Blue circles represent photon sites, orange squares represent qubit sites. On each node, the corresponding onsite energy is indicated. Nearest-neighbor photon sites are coupled to each other with coupling constant \(J\); the qubit sites are coupled to the corresponding photon sites with coupling constant \(g\).}
        \label{fig:hamilton_picture}
    \end{figure}
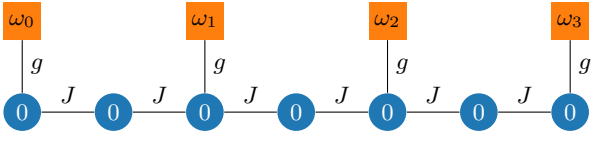

    We consider a one dimensional (1D) $N$-qubit wQED Hamiltonian \(H =  H_q + H_\gamma + H_{\gamma q}\) with terms corresponding to
	qubit energies \(H_q\), photon energies \(H_\gamma\) and coupling \(H_{\gamma q}\) between photons and qubits~\cite{caneva_quantum_2015}. As the experiment is performed in the low-power regime, interactions do not play a role and we limit ourselves to a non-interacting model. We therefore write for the qubits \(H_q / \hbar = \sum_{j=0}^{N-1} \omega_j a_j^\dagger a_j\) with the qubit creation and annihilation operators in site space \(a_j^\dagger\) and \(a_j\). The qubit energies \(\omega_j\) are drawn from a random Gaussian distribution \(P(\omega)=(2 \pi \sigma_{\omega}^2)^{-1/2} \exp[(\omega-\mu)^2/(2\sigma_{\omega}^2)]\) with standard deviation \(\sigma_{\omega}\) and mean \(\mu\). For the photon we choose a periodic dispersion relation \(H_\gamma / \hbar= -2J \sum_{n=0}^{N_{\gamma}-1} \cos(d_{\gamma} k_n) \tilde{b}^\dagger_{k_n} \tilde{b}_{k_n}\) with bandwidth \(4J\) and lattice spacing \(d_{\gamma}\); \(\tilde{b}^\dagger_{k}\) and \(\tilde{b}_k\) are photon creation- and annihilation operators in momentum space. Lastly, we specify \(H_{\gamma q} / \hbar = g \sum_{j=0}^{N-1} \sum_{n=0}^{N_{\gamma}-1} \left( \mathrm{e}^{\mathrm{i} k_n d j} \tilde{b}_{k_n}^\dagger a_j + \mathrm{h.c.} \right)\) where \(g\) is the coupling strength. The site-space equivalent of this Hamiltonian reads
	\begin{align}
		\frac{H}{\hbar} = &\sum_{i=0}^{N-1} \omega_i a_i^\dagger a_i - J\sum_{i,j=0}^{N_\gamma - 1} \delta_{\langle i, j\rangle} b^\dagger_i b_{j} \nonumber \\
		&+ g \sum_{i=0}^{N-1} \left( b_{i d / d_{\gamma}}^\dagger a_i + a_i^\dagger b_{i d / d_{\gamma}} \right),
		\label{eq:full_hamilton_site_space}\\
		\delta_{\langle i, j\rangle} =& \begin{cases}
		    1\,, & \ i,\ j \text{ nearest neighbors;}\\
		    0\,, & \ \text{else,}
		\end{cases} \nonumber
	\end{align}
here \(b^\dagger_i\), \(b_i\) are the site-space photon creation- and annihilation operators,
$d$ is the spacing between qubits, and \(d / d_{\gamma}\) is an integer. Note the position indices \(x = i d / d_\gamma\) of the photon operators in the coupling term: two neighboring qubits are coupled to
photon sites that are separated by \(N_{\rm int} = d / d_\gamma - 1\) intermediate photon sites, see Fig.~\ref{fig:hamilton_picture}. 
Taking the limit \(J\rightarrow \infty\) and \(d_\gamma \rightarrow 0\) while keeping \(d\) constant, one can restore the actual linear photon dispersion by adapting the band slope in the frequency range of interest to the speed of light. In practice (see below), one does not need a large number $N_{\rm int}$ of intermediate sites in the simulations; already $N_{\rm int} = 1$ as in Fig.~\ref{fig:hamilton_picture} turns out to be sufficient.

	For the analysis of disorder-induced localization in the wQED system, it is instructive to integrate out the qubit degrees of freedom in the Hamiltonian~\eqref{eq:full_hamilton_site_space} (rather than the photons as frequently done in the literature~\cite{caneva_quantum_2015,lalumiere_input-output_2013,shi_wqed}; see the Supplementary Material~\cite{supplementary} for the corresponding photon-integrated effective model), to obtain an effective photonic Hamiltonian. This can be conveniently done by using the generating integral $I$ for the Hamiltonian~\eqref{eq:full_hamilton_site_space} with the corresponding action \(S(\omega)\):
	\begin{widetext}
	\begin{align}
		I&= \int \mathcal{D}\{a, a^*, b, b^*\} \exp(-S(\omega))= \int \mathcal{D}\{a, a^*, b, b^*\} \exp(-\sum_{i=0}^{N - 1} (\omega a_i^* a_i) - \sum_{i=0}^{N_\gamma - 1}(\omega b_i^* b_i) + H) \nonumber \\
		&=\int \mathcal{D}\{a, a^*, b, b^*\} \exp(-\sum_{i, j=0}^{N -1 } a^*_i (\omega - \omega_i) \delta_{i, j} a_j + \sum_{i=0}^{N - 1} g( a_i^* b_{id / d_{\gamma}} +  b_{id / d_{\gamma}}^* a_i))\exp(\sum_{i, j=0}^{N_\gamma - 1} b_i^* (-J\delta_{\langle i, j \rangle} - \omega \delta_{i, j})b_j) \nonumber \\
		&\sim \int \mathcal{D}b \mathcal{D}b^* \exp(-\sum_{i, j=0}^{N_\gamma - 1} J b_i^* b_j \delta_{\langle i, j \rangle} + \sum_{i = 0}^{N- 1} b^*_{i d / d_{\gamma}} b_{i d / d_{\gamma}} \left[\frac{g^2}{\omega_i - \omega} - \omega\right]). \label{eq:I}
	\end{align}
	\end{widetext}
	Using the last line of Eq.~\eqref{eq:I}, we read off the effective photonic Hamiltonian:\newpage
	\begin{align}
		H_{\rm eff}(\omega)/\hbar &=\sum_{i=0}^{N - 1} n_{i(N_{\rm int} + 1)}  \frac{g^2}{\omega_i - \omega} -   J \sum_{i,j=0}^{N_\gamma - 1} \delta_{\langle i, j \rangle} b_i^\dagger b_j, \label{eq:h_eff_photon}\\
		n_i &= b_i^\dagger b_i \,, \nonumber
	\end{align}
	which depends on the considered frequency \(\omega\). For any fixed \(\omega\), we obtain a 1D Hamiltonian with onsite disorder corresponding therefore to an Anderson-localized system~\cite{scaling_localization}. (See Appendix~\ref{sec:weak_do_limit} for a discussion of the weak-disorder limit of the model.) 
	
	Due to the tridiagonal shape of $H_{\rm eff}$, we can efficiently calculate the average inverse localization length \(\frac{1}{\xi}\) \cite{paper:transfer_matrix} and the power transmission coefficient \(T\) \cite{paper:transmission} by solving the discrete Schr\"odinger equation with standard transfer-matrix methods (see Supplementary Material~\cite{supplementary}).
	Since localization implies exponential suppression of the wave function with system size, the localization length and the transmission are related as 
	\begin{align}
		\frac{1}{\xi} = \lim_{N\rightarrow\infty}  \left\langle - \frac{\log(T)}{N}\right\rangle =: \lim_{N\rightarrow\infty} \frac{1}{\xi_N}. \label{eq:transmission_localization_length}
	\end{align}
	Here and in the following, the localization length is given in units of the lattice constant; angular brackets denote an average over disorder realizations. With Eq.~\eqref{eq:transmission_localization_length}, effective localization lengths \(\xi_N\)  can be obtained from transmission through systems of any size. However, for small \(N\), the localization length \(\xi_N\) may be substantially influenced by finite-size effects. The limit \(N\rightarrow\infty\) implies that the system has to be sufficiently large for \(\xi_N\) to converge to the localization length $\xi$. \par
	
	In order to describe the experiment with the model, we need to find the appropriate parameters \(d\), \(\mu\), \(\sigma_{\omega}\), \(J\), \(d_{\gamma}\) and \(N_{\rm int}\). In accordance with the specifications of the sample the experiment is performed on (see Sec.~\ref{sec:experiment}), we set the distance between two qubits to \(d = \unit[400]{\mu m}\) and choose \(\mu / (2 \pi) = \unit[7.835]{GHz}\) for the center frequency. We also use the same values of disorder strength \(\sigma_\omega\) as in the experiment. (The disorder strength can be controllably varied in the experiment.) Furthermore, we can express the single-qubit 
	% radiative decay rate of the first excited state to the ground state
	decay width through the model parameters as \(\Gamma_{10} = g^2 / J\) (assuming \(\Gamma_{10} / J \ll 1\)). In correspondence to the experiment, we set \(\Gamma_{10} / (2 \pi) = \unit[6.4]{MHz}\). The parameters \(J\) and \(d_\gamma\) do not have direct equivalents in the experiment since they were introduced in the model by replacing the actual photon dispersion with a periodic spectrum. The linear dispersion is restored by considering a frequency window much smaller than \(J\) near the middle of the photon band. Near the middle of the band the effective photonic speed in the model is \(c = 2 J d_\gamma = 2J d / (N_{\rm int} + 1)\) and we calculate \(J\) from \(d\) and the speed of light in the sample, \(c = \unit[1.8\cdot 10^8]{m / s}\). 
	Lastly, we need to choose the number \(N_{\rm int}\) of the intermediate photon sites. Due to the photon dispersion in the lattice model, the physics is different at even / odd \(N_{\rm int}\) at the considered frequencies as the even case leads to an additional, frequency-independent phase shift between qubit-coupled sites, not corresponding to the linear photon dispersion. This is because linearization in the middle of the band gives us an approximate dispersion of \(\omega(k) \approx k c - J \pi\), and the corresponding phase shift \(\Delta \varphi = d k \approx \omega d / c +  (N_{\rm int} + 1) \pi/2\) between two qubit-coupled sites.  We observe that for even $N_{\rm int}$ there is an additional contribution $\pi$ to the phase shift, which makes possible spurious Fabry-P\'{e}rot resonances between neighbouring qubit sites. Thus, even $N_{\rm int}$ is less appropriate for modeling the photonic continuum, and we choose $N_{\rm int}$ to be odd. 
	While formally we need the limit of large \(N_{\rm int}\) to get a strictly linear dispersion, it turns out that 
	 \(N_{\rm int} = 1\) is fully sufficient for our purposes. This is because we work in a relatively narrow window around the center frequency \(\mu / (2\pi)\), in which the dispersion is approximately linear already for \(N_{\rm int} = 1\). (To verify  this, we have performed numerical calculations also for larger odd values of $N_{\rm int}$ and found no substantial difference.) 
	 We thus fix \(N_{\rm int} = 1\) in the following.
	 Using the experimental values of $c$, $d$, and $\Gamma_{10}$, we then find \(J=\unit[4.5 \cdot 10^{11}]{1/s} \) and \(g \approx \unit[4.25\cdot 10^9]{1 / s}\) from the above considerations.  Further, we introduce the dimensionless disorder strength via \(W = \sigma_{\omega} / \Gamma_{10}\).

As we show below, the model that we consider captures very well key physical properties	of our experimental setup. 
At the same time, this model involves an idealization of the experimental setup in several respects. In particular, the following additional effects present in the experiment are not taken into account by the model:
	\begin{itemize}
	    \item Experimental transmission spectra are influenced by interference effects in the cryostat~\cite{brehm_waveguide_2020}.
	    \item Non-radiative decays take place in the qubit array, introducing an additional suppression of transmission.
	\end{itemize}
	These effects are discussed in more detail in the next section.
	
	While our experimental setup contains a modest number of qubits, $N=8$, we study first the model in the limit of large $N$ that guarantee the convergence (i.e., essentially the limit $N\to\infty$). Later, we will analyze the impact of finite-size effects by comparing the $N \to  \infty$ and $N=8$ systems.
	
\begin{figure}[tb]
		\centering
		\includegraphics[width=\columnwidth]{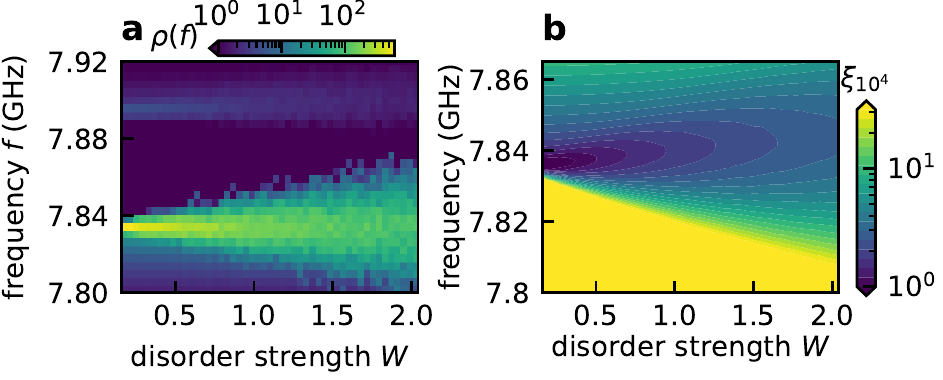}
		\includegraphics[width=\columnwidth]{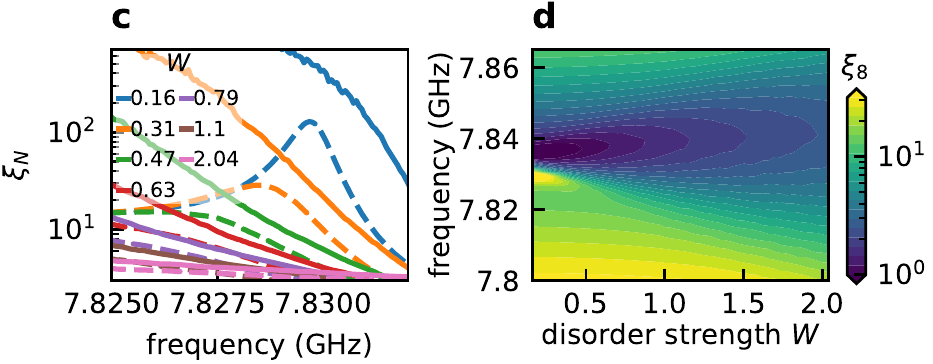} %,valign=b
		\caption{Model results for parameters \(J=\unit[4.5 \cdot 10^{11}]{1/s} \), \(g \approx \unit[4.25\cdot 10^9]{1 / s}\), and \(\mu /(2\pi) = \unit[7.835]{GHz}\) corresponding to the experimental values. \textbf{a} Average density of states (DOS) \(\rho(f)\) of Hamiltonian~\eqref{eq:full_hamilton_site_space} with \(N=2000\) and \(W\in [0.16, 2.04]\), for \(f \in [\unit[7.8]{GHz}, \unit[7.92]{GHz}]\). \textbf{b} Average localization length as a function of frequency and disorder strength \(W\) obtained in systems of \(N=10^4\) sites, averaged over \(40\) disorder realizations.  \textbf{c} Comparison of \(\xi_8\) as defined in Eq.~\eqref{eq:transmission_localization_length} (dashed lines, averaged over \(10^4\) disorder realizations) to \(\xi_{2000}\) (solid lines, averaged over \(10^2\) disorder realizations). \textbf{d} Average localization length \(\xi_8\) as a function of frequency and disorder strength \(W\) averaged over \(10^4\) disorder realizations.}
		\label{fig:loc_length_exp_parameters}	
	\end{figure}

	In Fig.~\ref{fig:loc_length_exp_parameters}a the average density of states \(\rho(f)\) of the Hamiltonian~\eqref{eq:full_hamilton_site_space} with \(N=2000\) is shown in the disorder range \(W\in [0.16, 2.04]\). At low disorder, the system exhibits a band gap of width \(\Delta f  \sim \unit[60]{MHz}\) around \(\mu / (2 \pi)\). The spectrum is asymmetric with most of the eigenvalues situated below the gap. Inside the gap, for \(f\gtrsim \unit[7.84]{GHz}\) and weak disorder, the density of states is very low. In this region, strongly suppressed transmission coefficients and correspondingly short  localization lengths are therefore expected.
	
	Values of the localization length obtained for frequencies in the lower band and in the gap region  are shown for disorder strengths \(W \leq 2.04\) in Fig.~\ref{fig:loc_length_exp_parameters}b.
	Here we used a very large system size, $N=10^4$, which ensures a precise determination of the displayed localization length $\xi \ll N$.
	Depending on the frequency $f$, we observe three qualitatively different regimes, regarding the localization length as a function of disorder:
	\begin{itemize}
    	\item At ``low'' frequencies \(f \lesssim \unit[7.83]{GHz}\), the localization length decreases with increasing disorder strength. This is the usual behaviour for an Anderson-localized system. In this regime, we predict the experimental observation of Anderson localization. We expect the experimentally determined localization lengths to be most accurate at sufficiently strong disorder, such that \(\xi \lesssim 8\). This minimizes  finite-size effects in our sample of size \(N=8\). The yellow area in the lower left corner of Fig.~\ref{fig:loc_length_exp_parameters}b corresponds to localization lengths much larger than the experimentally realized system size.
    	\item At ``high'' frequencies \(f \gtrsim \unit[7.845]{GHz}\), the localization length increases with increasing disorder strength---in contrast to the usual behaviour for disorder-induced localization. This can be understood by the dilution of the gap due to stronger disorder (cf. Fig.~\ref{fig:loc_length_exp_parameters}a). Another way to see this is to consider the onsite disorder in the effective model \(\varepsilon_i = \frac{g^2}{\omega - \omega_i}\): For sufficiently large \(\sigma_{\omega} \gg (\omega - \mu)\) an increase of the disorder strength \(\sigma_\omega\) leads to decreasing effective disorder values \(\varepsilon_i\).
    	\item In between, at frequencies \(\unit[7.83]{GHz} \lesssim f\lesssim \unit[7.845]{GHz}\), the system exhibits a crossover between the aforementioned regimes. Consequently, the localization length behaves non-monotonically as a function of disorder strength: it first decreases and then increases with increasing $W$. This can best seen at \(f \approx \unit[7.835]{GHz}\) where the same contour line is crossed twice when the disorder increases at constant frequency.
	\end{itemize}
    
	Having determined the localization length as a function of disorder strength and frequency, we can investigate how much of the localization physics is already visible in a disordered wQED system of experimentally realizable size. To this end, we calculate the power transmission coefficient \(T\) and the corresponding localization length \(\xi_N\) in a system with relatively small number $N$ of qubits and compare $\xi_N$  to \(\xi\).
	Specifically, we choose \(N=8\), corresponding to the number of qubits in the experimental sample (see below).
	At zero disorder, the transmission coefficient \(T(\omega)\) obtained from the Hamiltonian~\eqref{eq:h_eff_photon} features several peaks of perfect transmission, with strong reflection between the peaks. These peaks correspond to Fano-type resonances related to ``subradiant'', or ``dark'', modes observed in the experiment~\cite{brehm_waveguide_2020}. 
	Consideration of the clean system away from the peaks demonstrates the difference between
	the $N=8$ and \(N\rightarrow\infty\) models: while the $N=8$ system exhibits substantial reflection there, the infinite system is perfectly transmitting. 
	On the other hand, in the peak region, the small system mimics the behaviour of the infinite one, featuring a diverging localization length with \(T\rightarrow 1\) within numerical precision. With increasing system size the number of peaks would increase, gradually forming continuous bands corresponding to the converged localization length.
	
	We can now test, 
	focusing on the frequency range within the peaks (``dark modes''),
	how the introduction of disorder influences their behaviour. 
	The localization length $\xi_8$ in the experimentally relevant range of disorder and frequency is shown in Fig.~\ref{fig:loc_length_exp_parameters}c, in comparison with the thermodynamic-limit localization length $\xi$.
It can be seen that	\(\xi_8(\omega)\) at small disorder \(W=0.16\) features a peak at \(f_{\rm peak} \approx \unit[7.829]{GHz}\), which corresponds to a perfect transmission frequency at zero disorder. With increasing disorder strength, the peak frequency decreases and the peak smoothens out. While there is hardly any disorder dependence of the localization length away from the peaks, the peaks themselves get strongly suppressed with increasing disorder strength. This demonstrates that even a shorter finite system ($N=8$) shows indeed at the peak frequencies the behaviour characteristic for a large system.
In Fig.~\ref{fig:loc_length_exp_parameters}d we show \(\xi_8\) as a function of disorder strength and frequency, to be compared with \(\xi\) in Fig.~\ref{fig:loc_length_exp_parameters}b. Again, it can be seen that a small system indeed captures the behaviour of the large system qualitatively as well as quantitatively as long as the localization length is not much larger than the system size. In particular, all three qualitatively different types of behaviour of the localization length with increasing disorder strength are reproduced. As expected, significant deviations are present in the parameter regime of the yellow area in Fig.~\ref{fig:loc_length_exp_parameters}b, where \(\xi > 160 \gg 8\).
	
\section{Experiment}
\label{sec:experiment}

\begin{figure}[tb]
	\includegraphics[width=\columnwidth]{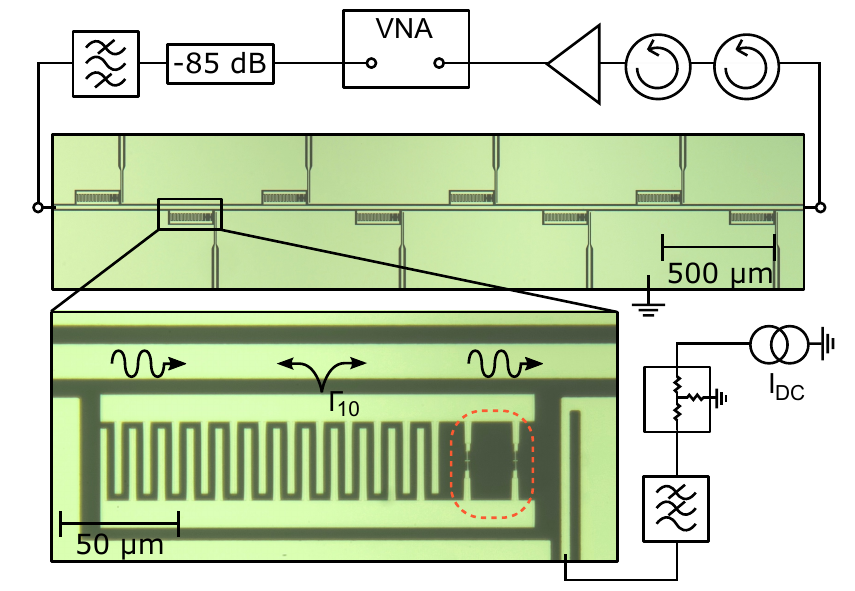}
	\caption{Microscopic image of the qubit metamaterial and sketch of the used measurement and control electronics. The metamaterial consists of eight superconducting transmon qubits coupled to a coplanar waveguide. All structures are fabricated from aluminum (bright areas) on a sapphire substrate (dark). Every qubit can be individually tuned with a local dc-flux bias line inducing magnetic flux in the qubit's split Josephson junction (red dotted).}
	\label{fig:Fig2}
\end{figure}

The localization is experimentally tested in a superconducting wQED system comprising eight transmon qubits~\cite{koch_charge-insensitive_2007}, coupled to a common coplanar waveguide (see Fig.~\ref{fig:Fig2}). The distance between neighbouring qubits is $d=400\,\upmu$m, which is at all accessible frequencies smaller than the corresponding wavelength $\lambda$. Each qubit is individually frequency tunable between 3 and $8\,$GHz with a local flux-bias line, which induces a magnetic flux in its SQUID. A calibration scheme is used to compensate for magnetic crosstalk of the bias lines, which allows for true individual qubit frequency control \cite{yang_probing_2020,brehm_waveguide_2020}. A schematic of this setup is provided in the Supplementary Material~\cite{supplementary}.

By measuring the transmission signal of the resonance fluorescence of the individual qubits, we extract the average radiative relaxation rate of $\Gamma_{10}/(2\pi)=6.4\,$MHz and an average decoherence rate of $\gamma_{10}/(2\pi)=(\Gamma_{10}/2+\Gamma_\text{nr})/(2\pi)=3.6\,$MHz around the qubit frequencies of $7.9\,$GHz \cite{brehm_waveguide_2020}. The average non-radiative decoherence rate is $\Gamma_\text{nr}/(2\pi)=400\,$kHz. The incident microwave power is kept far below the single photon regime ($P<\hbar\omega_\text{c}\Gamma_{10}$). This avoids saturation of the qubits and eliminates the interaction in the system, which allows to study the single-particle regime where Anderson localization is expected.

%%%%%%%%%%%%%%
%%%%%%%%%%%%%%%

We introduce disorder to the system by tuning the qubits to random frequencies $\omega_i$ picked according to a Gaussian probability distribution $P(\omega)$ with standard deviation $\sigma_{\omega}$. The distribution is centered around $\mu/(2\pi)=7.835\,$GHz, which is near the qubits' upper flux sweet spot. The interval of allowed qubit frequencies is limited to $(\omega_i-\mu) \in \left[-2.5\cdot \sigma_{\omega}, +2.5\cdot \sigma_{\omega}\right]$. Again, the disorder strength is characterized by the dimensionless variable $W=\sigma_{\omega}/\Gamma_{10}$. All spectroscopic measurement data in this work is normalized with $S^\text{norm}_{21}(\omega)=S^\text{meas}_{21}(\omega)/\text{max}(S^\text{meas}_{21}(\omega))$ where \(S_{21}\) is the transmission coefficient.

\begin{figure}[tb]
	\includegraphics[width=\columnwidth]{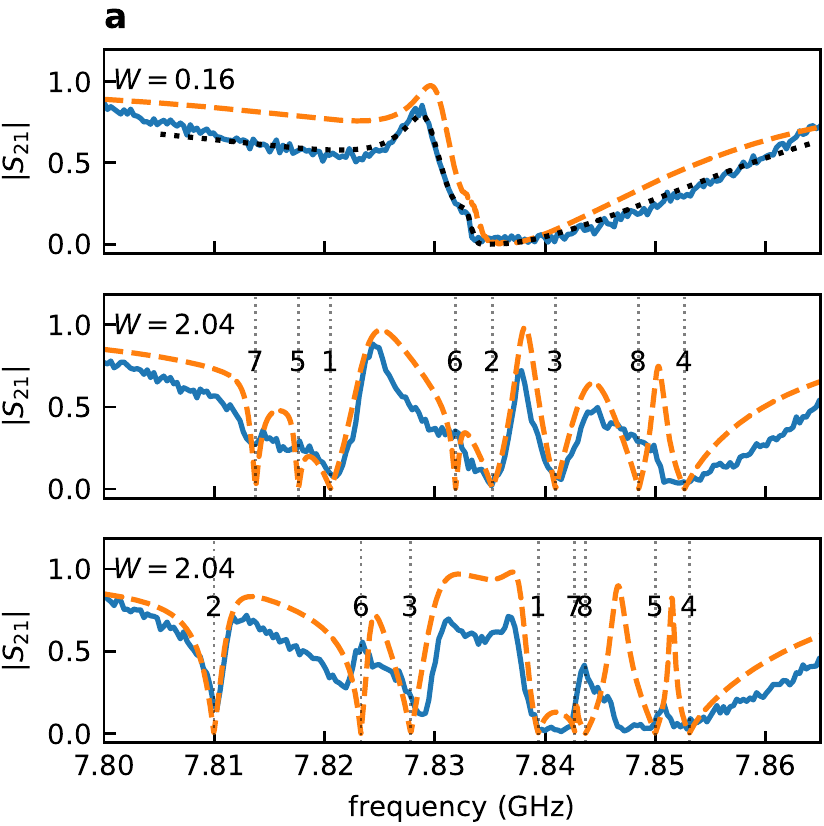}
	\includegraphics[width=\columnwidth]{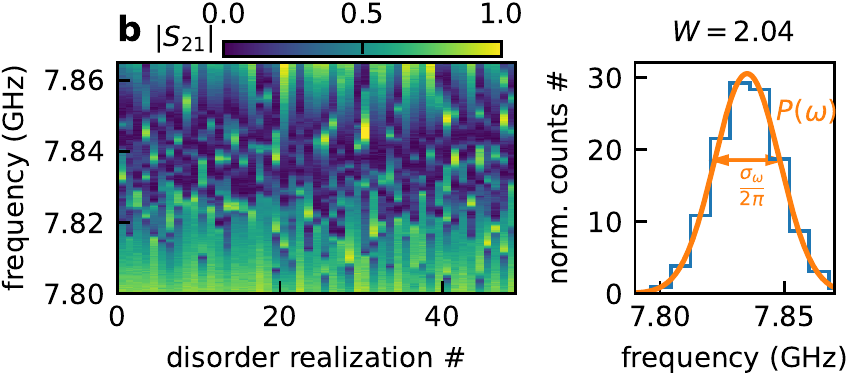}
	\caption{\textbf{a} Transmission coefficient \(|S_{21}|\) through the metamaterial for different disorder strengths $W$ in the experiment (blue solid lines) and in the theoretical model (orange dashed lines). For the case of very weak disorder, $W=0.16$ (upper panel), the observed transmission is nearly identical to the results of a transfer-matrix calculation for a clean system. The agreement becomes especially good when the non-radiative decay in the sample as well as the background in the cryostat are incorporated (black dotted line). For strong disorder, $W=2.04$, the middle and the bottom panels display results for two individual disorder realizations. Vertical dotted lines mark the frequency of the individual qubits in each disorder realization.
	It is seen that the profile of the transmission coefficient differs strongly from one realizations of disorder to the other, in agreement with theoretical predictions.
	\textbf{b} Overview of the transmission coefficient for 50 random disorder realizations for $W=2.04$. The distribution of qubit frequencies is picked according to a Gaussian probability distribution $P(\omega)$ (right panel).}
	\label{fig:Fig3}
\end{figure}

We first consider the transmission $|S_{21}|$ through the sample at very weak disorder $W=0.16$ (Fig.~\ref{fig:Fig3}a, top panel), where it is similar to the clean case, which is discussed in detail in Ref.~\cite{brehm_waveguide_2020}. In the experimental data (blue solid line), only the brightest of the subradiant modes is visible as a pronounced peak. The experimental results compare very well with  theoretical data obtained by using the model of Sec.~\ref{sec:theory} (orange dashed line). 
%\sout{There are small deviations of the theoretical curve from the experimental one, which include a slightly larger transmission amplitude at the brightest mode as well as a small additional peak at \(f\approx \unit[7.832]{GHz}\).} 
There are small deviations of the theoretical curve from the experimental one; in particular, the theoretical transmission amplitude at the peak is slightly larger. As was already mentioned in 
Sec.~\ref{sec:theory}, these differences can be explained by the non-radiative decay in the sample as well as by the background in the cryostat. We confirm this by fitting the experimental data to a transfer-matrix calculation (black dotted line) incorporating both effects on a phenomenological level (for details of the procedure see Ref.~\cite{brehm_waveguide_2020}). The asymmetric line shape of the transmission dip is a manifestation of the asymmetric opening of the band gap (for $\omega>\mu$) as seen from the theoretical calculation, Fig.~\ref{fig:loc_length_exp_parameters}a.

For larger disorder strengths, the pronounced peak of the collective eight-qubit subradiant state is not observable anymore. Instead, the line shape changes to an ensemble of peaks and dips, which differs significantly between different disorder realizations (Fig.~\ref{fig:Fig3}a, b) and is well reproduced by the model (orange dashed lines). In Fig.~\ref{fig:Fig3}a, the individual qubit frequencies are indicated by grey dotted lines. It is seen, both in theoretical and in experimental curves, that every frequency corresponds to a distinct dip in the transmission. Recalling our convention for the disorder strength, \(W = \sigma_{\omega} / \Gamma_{10}\), this can be intuitively understood from the fact that the difference between two qubit frequencies is typically larger than the single-qubit decay width if \(W \gtrsim 1\).

\begin{figure}[tb]
	\includegraphics[width=\columnwidth]{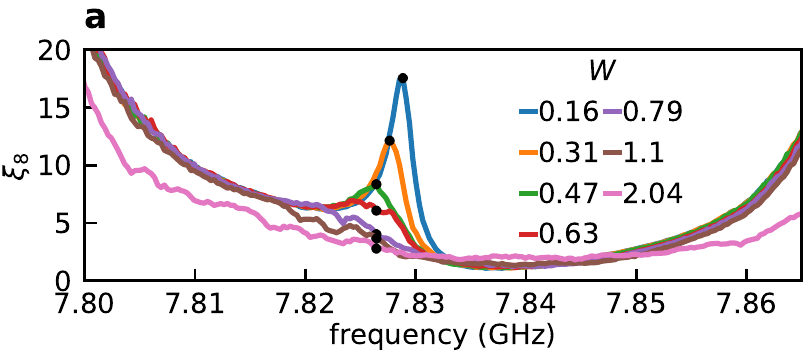}
	\includegraphics[width=\columnwidth]{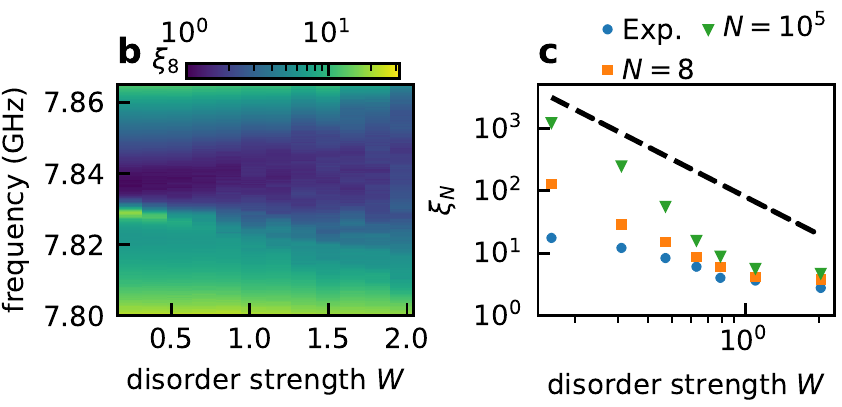}
	\caption{\textbf{a} Localization length $\xi_8$ obtained experimentally from an average of 50 random disorder realizations. Different curves correspond to different values of the disorder strength $W$. Large localization lengths at the peak of the brightest collective subradiant mode are suppressed with increasing disorder strength, which demonstrates the Anderson localization of waveguide photons. \textbf{b} Dependence of the measured localization length $\xi_8$ on frequency and disorder strength $W$. Each vertical trace is based on an average of 50 random disorder realizations. \textbf{c} Colored symbols: Disorder-strength dependence of the measured localization length at the peak of the brightest subradiant state (marked as blacked dots in the panel \textbf{a}), in comparison with the theoretical calculation for a small ($N=8$) and a large ($N=10^5$) system. The theoretical localization lengths at \(W \in \{0.16, 0.47\}\) were calculated at slightly different frequencies since the peak positions differ by a small amount of approximately \(\unit[1]{MHz}\). Dashed black line: Power law \(\xi \sim W^{-2}\) as a guide to the eye.}
	\label{fig:Fig4}
\end{figure}

In order to analyze the localization length in the sample, the disorder strength is varied between $W=0.16$ and $W=2.04$ and, for each value of $W$, the transmission $|S_{21}|$ of fifty random disorder realizations is measured. Using Eq.~\eqref{eq:transmission_localization_length}, we obtain $\xi_8$ from the logarithmic averaged power transmission $T=|S_{21}|^2$. The resulting dependencies of the localization length $\xi_8$ on frequency are shown in Fig.~\ref{fig:Fig4}a for several disorder strengths. In Fig.~\ref{fig:Fig4}b, the same experimental data for $\xi_8$ are presented by a color code in the parameter plane spanned 
by disorder strength and frequency.
These experimental findings are in good agreement with the corresponding theoretical results in Fig.~\ref{fig:loc_length_exp_parameters}c (dashed lines) and Fig.~\ref{fig:loc_length_exp_parameters}d. For a direct comparison of the frequency dependence of the localization length $\xi_8$ in theory and experiment for each disorder strength, see Appendix~\ref{sec:comparison_of_av_loc_lengths}. As in the theoretical model, we find in the experiment a strong reduction of $\xi_8$ at the peak of the brightest subradiant mode around $f \approx 7.829\,$GHz with increasing disorder strength $W$. For the largest disorder strength of $W=2.04$, the localization length is reduced to $\xi_8\approx 3$, implying that the localization is strong in comparison with the overall length of $N=8$ unit cells of the structure. Comparing Figs.~\ref{fig:loc_length_exp_parameters}b and \ref{fig:Fig4}b, we find all three qualitatively different regimes predicted by the theory reproduced in the experiment: For \(f\lesssim \unit[7.82]{GHz}\), the localization length decreases with increasing disorder strength, showing disorder-induced localization. For \(f\gtrsim\unit[7.84]{GHz}\), the localization length increases with increasing disorder strength due to the dilution of the band gap. In the crossover region in between, the localization length behaves non-monotonically as a function of disorder.

In Fig.~\ref{fig:Fig4}c we show the disorder dependence of the localization length at the peak, as obtained experimentally, and compare it to the theoretical calculations for $N=8$ and in the large-$N$ limit. (Note that the peak frequencies in model and experiment exhibit a small difference of approximately  \(\unit[1]{MHz}\), so that we compare the localization lengths at slightly different frequencies for \(W \in\{0.16, 0.47\}\).) At weak disorder, the large-\(N\) theoretical data is close to a power law \(\xi\sim W^{-2}\) (black dashed line). This is expected from perturbative analysis and discussed in more detail in Appendix~\ref{sec:weak_do_limit}, where we also demonstrate an excellent $1/W^2$ scaling of the localization length by using weaker disorder and much larger system sizes. 

For the weakest disorder strength in Fig.~\ref{fig:Fig4}c, $W=0.16$, the experimentally determined localization length is significantly smaller than the actual \(\xi\), which is rooted in \(\xi \gg 8\) for these parameters. Furthermore, the experimentally extracted $\xi_8$ is also smaller by almost an order of magnitude than the theoretical \(\xi_8\), which is due to the above-mentioned additional peak-suppressing effects in the experiment, which are not present in the model. In particular, dissipation, i.e., non-radiative decay of the qubits, yields an additional contribution to $\Im(k)$ for frequencies below and above the bandgap \cite{Mirhosseini2018}. This effect reduces the maximum $T$ at the subradiant polariton peak and limits the experimentally extracted localization length from above, even in the absence of disorder.
A further potential source of differences between the theory and the experimental data is the interference between the signal and standing waves in the cryostat~\cite{brehm_waveguide_2020}. This effect leads, e.g., to an asymmetry in the one-qubit transmission dip~\cite{brehm_waveguide_2020} and accounts for further suppression of the localization length in experiment.
At the same time, as the disorder strength is increased, the disorder-induced peak suppression becomes the dominant effect.
Indeed, it is evident in Fig.~\ref{fig:Fig4}a that the peak height decreases with the increase of disorder.
On a quantitative level, all effective localization lengths (experimental $\xi_8$ as well as theoretical $\xi_8$ and $\xi$) converge towards each other. In particular, for 
\(W = \unit[1.1]{}\) the experimentally determined $\xi_8$ is smaller than the theoretical $\xi_8$ ($\xi$) by approximately \(\unit[10]{\%}\) (respectively, \(\unit[40]{\%}\)). 
Furthermore, we explicitly demonstrate  in Appendix~\ref{sec:diss_vs_loc}, by incorporating the non-radiative decay rate into the transfer-matrix approach, that the influence of dissipation associated with the non-radiative decay becomes negligible at \(W\gtrsim 1\). Since the experiment reaches sufficient disorder strengths and shows a distinct dependence on disorder, we conclude, in agreement with the results of the theoretical model, that the observed suppression of the subradiant mode peak is an observation of Anderson localization in an eight-qubit system.

%%%%%%%%%%%%%%%%%%%%%
%%%%%%%%%%%%%%%%%%%

%%%%%%%%%%%%%%%%%%%%
%%%%%%%%%%%%%%%%

\section{Summary and outlook}

In conclusion, we experimentally demonstrated Anderson localization in a finite wQED system with $N=8$ qubits. Individual frequency control allowed us to manipulate the disorder strength in-situ. Attributing an effective localization length to the suppression of a subradiant polaritonic mode, we found a decrease of the effective localization length with increasing disorder. Using a theoretical model, we showed that in finite systems localization indeed manifests as the suppression of transmission peaks, which constitute the polariton bands for infinite structures. The model allowed us to compare localization lengths obtained in small systems to the thermodynamic-limit ($N \to \infty$) results. We found that the finite system mimics the infinite one at the peak positions, giving a reasonable estimate for the localization length at the corresponding frequencies. 
The theoretical analysis reproduces all salient aspects of the experiment. This confirms the accuracy of the experiment and corroborates the appropriateness of the model as well the interpretation of the experimental findings.

 The experiment paves the way to further experimental studies in disordered wQED systems. One very interesting prospective direction is the implementation of an individual readout of the qubits. This will give access to dynamical properties of the system by studying observables such as the return probability or the wave packet spread. Such studies would also contribute to investigation of finite-size effects. Furthermore, an individual qubit readout would allow one to probe the spatial profile of localized collective excitations, providing a potential experimental platform to test the localization landscape technique \cite{landscape1,landscape2}. Finally, a very intriguing and challenging  prospect is to explore the regime of many-body localization by increasing the incident power in the system.

\section{Acknowledgements}
The authors thank A.N. Poddubny and J.F. Karcher for helpful discussions. J.D.B. acknowledges support from Studienstiftung des Deutschen Volkes. We acknowledge funding by Deutsche Forschungsgemeinschaft (DFG) via grants No. GO 1405/6-1 and US 18/15-1, and also support by the European Union’s Horizon 2020 Research and Innovation Programme under Grant Agreement No. 863313 (SUPERGALAX). A.V.U. acknowledges support of this work by the Russian Science Foundation via grant No. 21-72-30026. 
%partial support
%from the Ministry of Education and Science of %the Russian Federation in the
%framework of the Increase Competitiveness %Program of the National University of
%Science and Technology MISIS (contract No. K2-2020-022).

\clearpage
\appendix

\section{Localization vs. dissipation}
\label{sec:diss_vs_loc}

\begin{figure}[tb]
	\includegraphics[width=\columnwidth, trim={3cm 0 3cm 0}, clip]{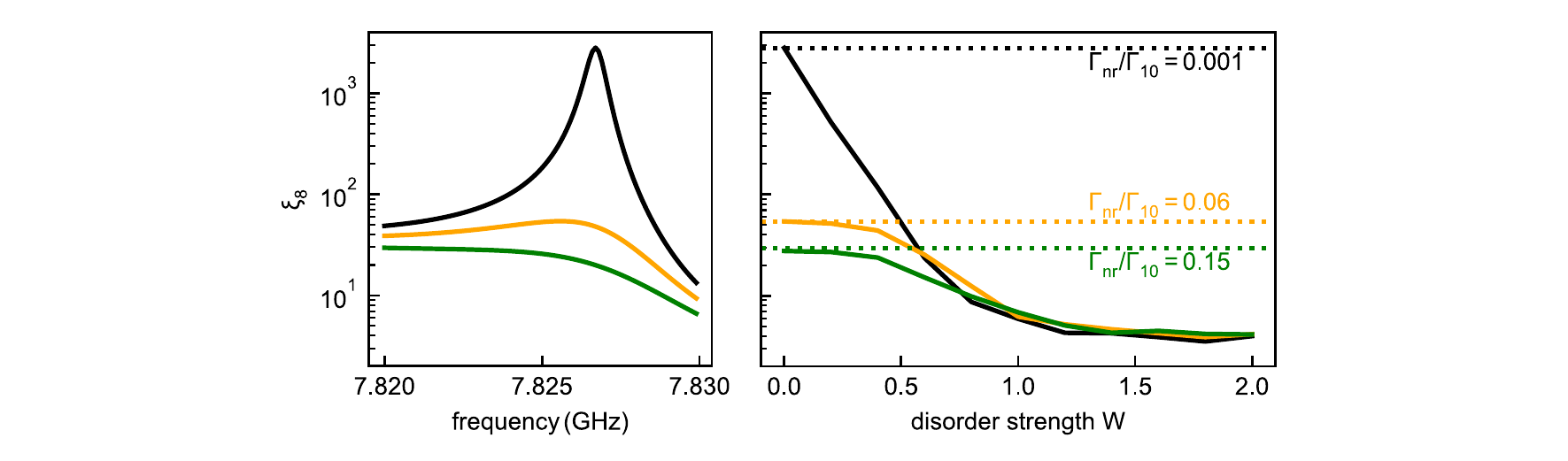}
	\caption{Effect of dissipation (non-radiative decay) on the effective localization length. Left panel: numerically calculated $\xi_8=- 8/\log(T)$ for the $N=8$ clean system obtained from a transfer-matrix calculation around the frequency of the brightest subradiant mode. Here, the dissipation is included via the non-radiative decoherence rate $\Gamma_\text{nr}$ of the qubits. With increasing dissipation, the value of $\xi_8$ at the peak gets progressively suppressed.
	%which has a comparable effect as the presence of disorder of reducing $\xi_8$ at the position of the peak.
	Right panel: $\xi_8=- 8/\left\langle\log(T)\right\rangle$ at the position of the peak as a function of disorder. Even in the presence of dissipation, $\xi_8$ monotonously decreases with increasing disorder due to Anderson localization. At sufficiently strong disorder, the dissipation does not play any essential role. }
	\label{fig:SupplFig3}
\end{figure}

Since the localization length is in the experiment extracted from the power transmission coefficient $T=|S_{21}|^2$, the effective localization length $\xi_8 = - \left\langle 8 / \log(T) \right\rangle$ is also influenced by dissipation (i.e. non-radiative decoherence) of the qubits. This can be numerically investigated with a phenomenological transfer-matrix approach based on the reflection coefficient of the individual qubits, which incorporates non-radiative decay. The calculation follows the methods presented in Refs.~\cite{brehm_waveguide_2020,garcia_2017}. The left panel of Fig.~\ref{fig:SupplFig3} shows the calculated $\xi_8=-8/\log(T)$ for the clean system (all qubits at $\omega_i/2\pi=7.835\,$GHz, $\Gamma_{10}/2\pi=6.4\,$MHz) and different non-radiative decoherence rates $\Gamma_\text{nr}$. Although the maximum value of $\xi_8$ at the Fano-peak of the subradiant mode is strongly influenced by the amount of dissipation, it shows a distinct monotonic decrease with the disorder strength $W$ (right panel of Fig.~\ref{fig:SupplFig3}), which is a manifestation of Anderson localization. For larger disorder strengths, $\xi_8$ converges to the value of a non-dissipative model. 

%%%%%%%%%%%%%%%%%%%%%%
%%%%%%%%%%%%%%%%%%%

\section{Comparison of averaged experimental and theoretical localization lengths $\xi_8$ for various disorder strengths}
\label{sec:comparison_of_av_loc_lengths}

\begin{figure}[tb]
	\includegraphics[width=\columnwidth]{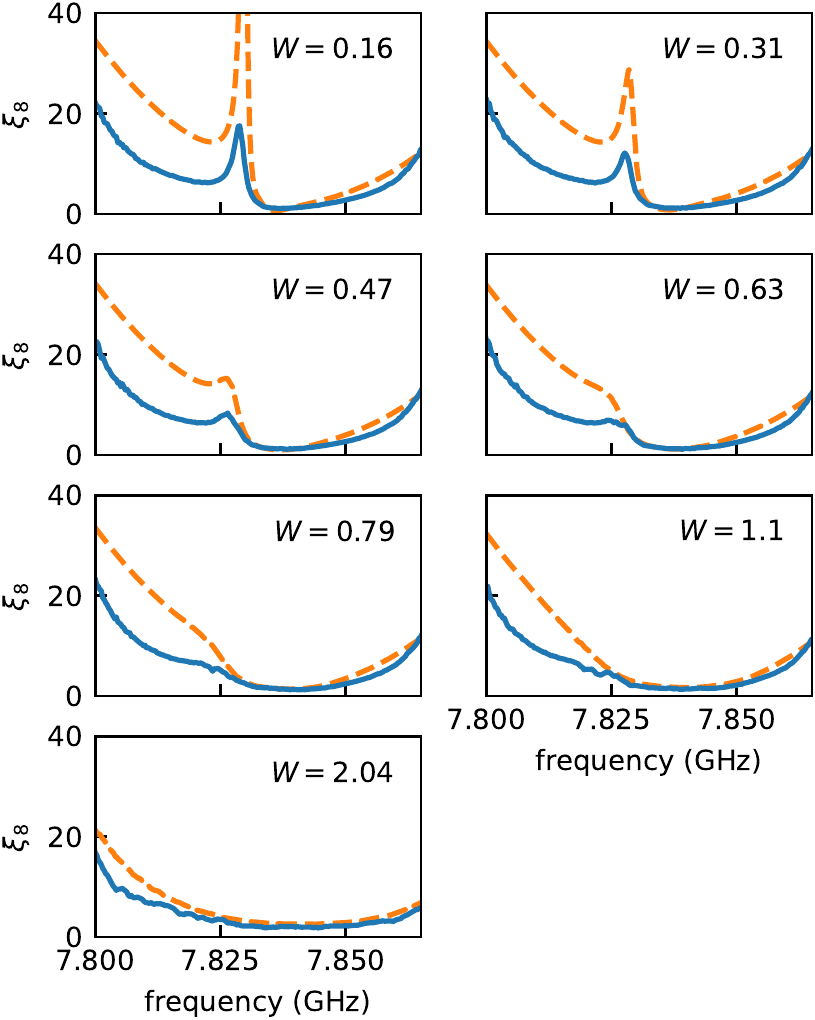}
	\caption{Comparison of frequency dependences of the localization length $\xi_8$ in experiment (solid blue lines) and theory (dashed orange lines) for different strengths of disorder \(W\in [0.16, 2.04]\). The experimental and theoretical data was averaged over 50 and \(10^4\) random disorder realizations, respectively.}
	\label{fig:comparison_average_loc_length}
\end{figure}

In Fig.~\ref{fig:comparison_average_loc_length} we show a direct comparison of the averaged localization lengths \(\xi_8\) in experiment (solid blue lines) and in the theoretical model (dashed orange lines) for various values of disorder $W$ in the range  \(W\in[0.16, 2.04]\). There is good agreement between the experiment and the theory with respect to the overall behavior of the localization length. At the same time, the experimental $\xi_8$ is substantially lower than the theoretical one for weak disorder, in the range of frequencies where the theoretical $\xi_8$ is large (in particular, at the peak). The reasons for this are discussed in Sections \ref{sec:theory} and \ref{sec:experiment} of the paper: the experimental data are affected by the non-radiative decay and by effects in the cryostat that are not included in the theoretical model. 
With increasing disorder, the agreement between the experimental and theoretical values of $\xi_8$ improves, as Anderson localization becomes the most important effect controlling the experimentally determined localization length.

%%%%%%%%%%%%%%%%%
%%%%%%%%%%%%

\section{Model \eqref{eq:full_hamilton_site_space} in the weak disorder limit}
\label{sec:weak_do_limit}
\begin{figure}[tb]
	\includegraphics[width=\columnwidth]{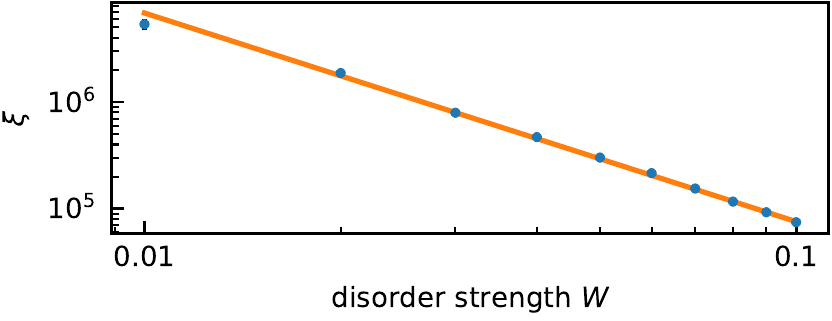}
	\caption{Localization length $\xi$ in the model \eqref{eq:full_hamilton_site_space} at frequency \(f=\unit[7.82]{GHz}\) and weak disorder, \(W\in[10^{-2}, 10^{-1}]\) determined for a system of size \(N = 3\cdot 10^7\) (blue dots). The data are based on averaging over 40 disorder realizations. The error bars were obtained with a bootstrapping procedure. The orange line is a power-law fit, \(\xi(W) \propto W^{-\beta}\) with \(\beta = 1.97\). The obtained value of $\beta$ is very close to the analytical prediction $\beta=2$ in the weak-disorder limit.}
	\label{fig:weak_disorder_limit}
\end{figure}

For weak disorder, the localization length scales as  $\xi\propto W^{-\beta}$ with $\beta=2$. This is because the localization length in 1D geometry is given by the back-scattering mean free path, which scales as $1/W^2$ by virtue of Fermi's golden rule. To verify with a high accuracy numerically that our model fulfills this limit, we have performed numerical simulations in the range of very weak disorders, $0.01 \le W \le 0.1$ and for a very big system size \(N = 3\cdot 10^7\). The results are shown in Fig.~\ref{fig:weak_disorder_limit}. The power-law fit yields the exponent \(\beta \approx 1.97\), in excellent agreement with the analytical prediction $\beta=2$. 

We can estimate an upper border of the range of disorder strength $W$ in which this behavior can be reasonably observed. 
For this purpose, we use the effective photonic Hamiltonian \eqref{eq:h_eff_photon}. There,
the effective onsite random field reads
\begin{align}
    \varepsilon_i = \frac{g^2}{\omega - \omega_i}.
\end{align}
Investigating the model at a frequency 
$\omega = \mu + \Delta \omega$ and assuming
$\sigma_\omega / \Delta\omega \ll 1$, we find 
\begin{align}
    \varepsilon_i & \simeq \frac{g^2}{ \Delta \omega} + \frac{(\omega_i - \mu) g^2}{(\Delta \omega) ^2}.
\end{align}
Therefore, the characteristic magnitude of fluctuations of the random on-site field is $\sigma_{\rm eff} = \sigma_\omega g^2/ (\Delta \omega)^2$. The weak-disorder condition reads
\(\sigma_{\rm eff} / J \ll 1\). We further use $\Delta \omega \approx \Gamma_{10}$, as in the experiment, with \(\Gamma_{10} = g^2 / J\) being the single-qubit radiative decay rate. The weak-disorder condition then takes the form 
\begin{align}
    W \equiv \sigma_\omega / \Gamma_{10} \ll 1 \,.
    \label{eq:dis_strength_w_2}
\end{align}
Thus, we expect that the $1/W^2$ weak-disorder scaling of the localization length $\xi$ can be observed with reasonable accuracy up to $W \lesssim 1$. This is indeed what is seen in Fig.~\ref{fig:Fig4}c.

%%%%%%%%%%%%%%%%%
%%%%%%%%%%%%%%%%%%%%

\clearpage
\bibliography{bibliography}

\clearpage
\newpage

\setcounter{figure}{0}
\setcounter{equation}{0}
\renewcommand{\thefigure}{S\arabic{figure}}
\renewcommand{\thetable}{S\arabic{table}}
\renewcommand{\theequation}{S\arabic{equation}}

\title{Supplemental Material: Tunable Anderson Localization of Dark States}
\author{Jan David Brehm}
	\affiliation{Physikalisches Institut, Karlsruhe Institute of Technology, 76131 Karlsruhe, Germany}
\author{Paul Pöpperl}
	\affiliation{Institut für Theorie der Kondensierten Materie, Karlsruhe Institute of Technology, 76128 Karlsruhe, Germany}
	\affiliation{Institute for Quantum Materials and Technologies, Karlsruhe Institute of Technology, 76021 Karlsruhe, Germany}
\author{Alexander D. Mirlin}
	\affiliation{Institut für Theorie der Kondensierten Materie, Karlsruhe Institute of Technology, 76128 Karlsruhe, Germany}
	\affiliation{Institute for Quantum Materials and Technologies, Karlsruhe Institute of Technology, 76021 Karlsruhe, Germany}
	\affiliation{Landau Institute for Theoretical Physics, 119334 Moscow, Russia}
\author{Alexander Shnirman}	
    \affiliation{Institut für Theorie der Kondensierten Materie, Karlsruhe Institute of Technology, 76128 Karlsruhe, Germany}
    \affiliation{Institute for Quantum Materials and Technologies, Karlsruhe Institute of Technology, 76021 Karlsruhe, Germany}
\author{Alexander Stehli}
	\affiliation{Physikalisches Institut, Karlsruhe Institute of Technology, 76131 Karlsruhe, Germany}
\author{Hannes Rotzinger}
	\affiliation{Physikalisches Institut, Karlsruhe Institute of Technology, 76131 Karlsruhe, Germany}
		\affiliation{Institute for Quantum Materials and Technologies, Karlsruhe Institute of Technology, 76021 Karlsruhe, Germany}
\author{Alexey V. Ustinov}
	\affiliation{Physikalisches Institut, Karlsruhe Institute of Technology, 76131 Karlsruhe, Germany}
		\affiliation{Institute for Quantum Materials and Technologies, Karlsruhe Institute of Technology, 76021 Karlsruhe, Germany}
	\affiliation{National University of Science and Technology MISIS, Moscow 119049, Russia}
	\affiliation{Russian Quantum Center, Skolkovo, Moscow 143025, Russia}

\date{\today}

\maketitle

\onecolumngrid

\section*{Experimental Setup}

\begin{figure}[tb]
	\includegraphics[width=\columnwidth]{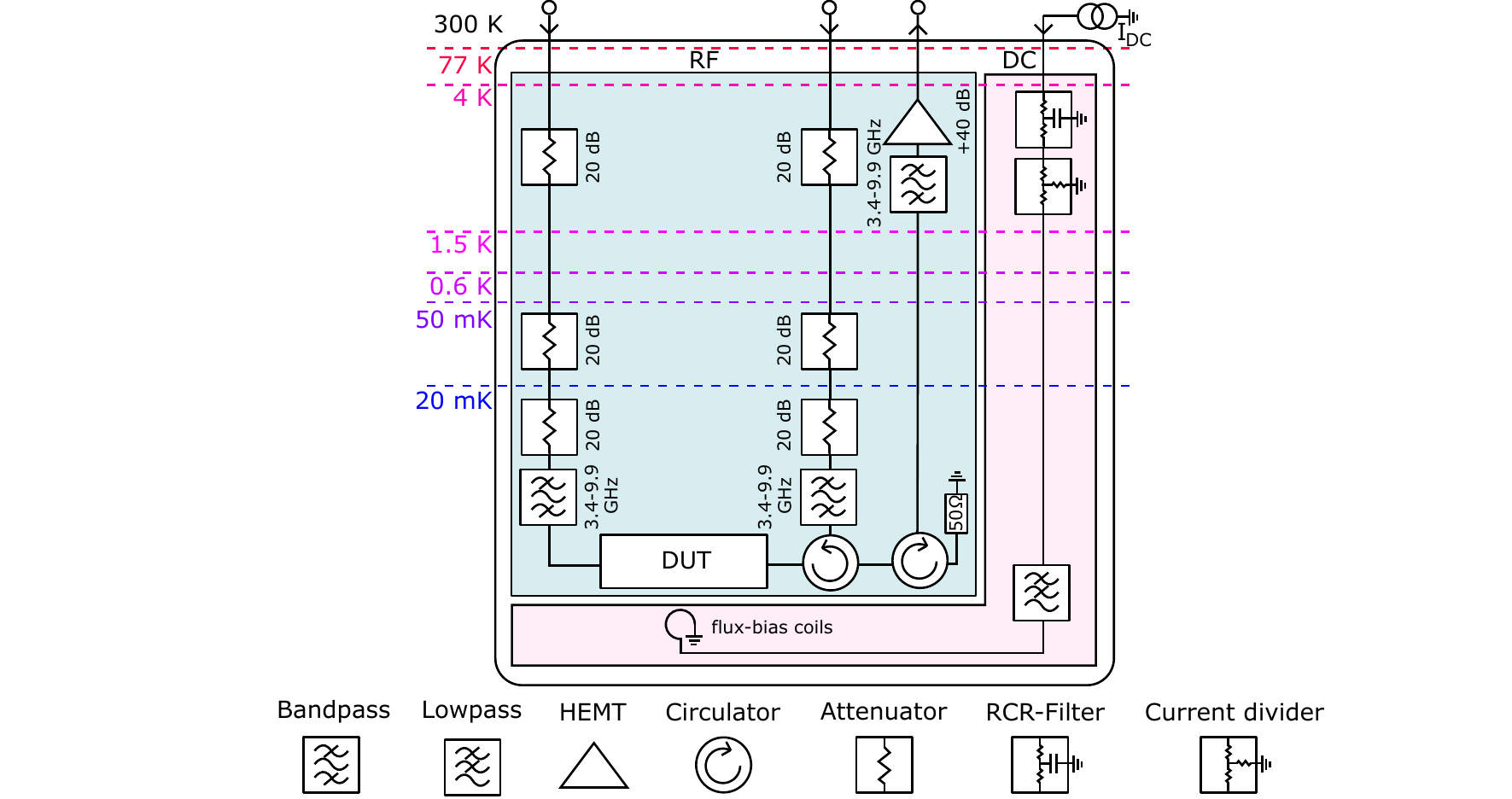}
	\caption{Sketch of the used cryogenic measurement setup.}
	\label{fig:SupplFig1}
\end{figure}

The cryogenic microwave measurement setup is shown in Fig.~\ref{fig:SupplFig1}.

%%%%%%%%%%%%%%
%%%%%%%%%%%%%

\section*{Effective Qubit Hamiltonian}
\label{sec:model_hamiltonian}

Our starting point in the theoretical part of the paper is the Hamiltonian (1) that describes an array of qubits coupled to a photonic waveguide. We proceed by integrating out qubit degrees of freedom, which leads to a photonic Hamiltonian (3), which is then analyzed via the transfer-matrix approach.

At the same time, a common course of action in the context of wQED systems 
is to integrate out the photonic degrees of freedom, which yields an effective qubit Hamiltonian
Refs.~\cite{lalumiere_input-output_2013, caneva_quantum_2015}. Since the resulting Hamiltonian is of long-range character, this approach is less convenient for the analysis of Anderson localization and is not used in our work. Nevertheless, we find it instructive to outline the derivation of the effective qubit Hamiltonian in this section of the Supplementary Material, in order to better clarify relations to previous works. 
We consider a Hamiltonian \(H = H_\gamma + H_q + H_{\gamma q}\) with terms corresponding to
	qubit part \(H_q\), photon part \(H_\gamma\) and coupling \(H_{\gamma q}\) between them: 
	\begin{align}
		H_q &= \sum_{j=1}^N \omega_j a_j^\dagger a_j \,, \\
		H_\gamma &= \sum_{n=1}^N \omega_{k_n} \tilde{b}^\dagger_{k_n} \tilde{b}_{k_n} \,, \label{eq:site_space_h_appendix}\\
		H_{\gamma q} &= g \sum_{j, n=1}^N \left( \exp(\mathrm{i} k_n d j) \tilde{b}_{k_n}^\dagger a_j + \exp(-\mathrm{i}k_n d j) a_j^\dagger \tilde{b}_{k_n} \right).
	\end{align}
    Here \(b^\dagger\), \(b\) and \(a^\dagger\), \(a\) are photon and qubit creation and annihilation operators, respectively, \(g\) is the coupling strength between photons and qubits, $\omega_i$ are random qubit frequencies, \(d\) the lattice spacing of the qubit array, and $N$ the number of qubits. We set \(\hbar\equiv 1\).
    
	Let us consider first the simplest lattice model, with the number of photon sites equal to the number of qubits $N$ (i.e., the photon lattice spacing equal to the qubit lattice spacing). The photon dispersion is then \(\omega_k =2 J \cos(dk)\). Transforming the photon operators from momentum space to the real space, we get 
	\begin{align}
		b^\dagger_j &:= \sum_{n=1}^N \exp(\mathrm{i}k_n d j) \tilde{b}^\dagger_{k_n} \,, \\
		H_\gamma &= J\sum_{i=1}^N \left( b^\dagger_i b_{i + 1} + b^\dagger_{i + 1} b_i \right) \,,\\
		H_{\gamma q} &= g \sum_{i=1}^N \left( b_{i}^\dagger a_i + a_i^\dagger b_i \right).
	\end{align}
	Therefore, we have the site-space Hamiltonian
	\begin{align}
		H &= \sum_{j=1}^N \omega_j a_j^\dagger a_j + J\sum_{i=1}^N \delta_{\langle i, j\rangle} b^\dagger_{i} b_j + g \sum_{i=1}^N \left( b_{i}^\dagger a_i + a_i^\dagger b_i \right);
		\label{eq:full_hamilton_site_space_app}\\
		\delta_{\langle i, j \rangle} &:= \begin{cases}
		    1 & i,\ j \text{ nearest neighbors}\\
		    0 & \text{else}.
		\end{cases}
	\end{align}
 Writing down a functional integral corresponding to this Hamiltonian and integrating out the photons, we obtain
	\begin{align}
		I &= \int \mathcal{D}\{a, a^*, b, b^*\} \exp(-\sum_{i, j=1}^N a^*_i (\omega - \omega_i) \delta_{i, j} a_j + g \sum_{i=1}^N (a_i^* b_i +  b_i^* a_i))\exp(\sum_{i, j=1}^N b_i^* (J\delta_{\langle i, j \rangle} - \omega \delta_{i, j})b_j)\\
		&=  \int \mathcal{D}\{a, a^*, b, b^*\} \mathrm{e}^{-S_a}\times \nonumber \\
		&\times g\exp(g \sum_{i, n=1}^N ( \exp(\mathrm{i} d i k_n) a_i^* \tilde{b}_{k_n} + \exp(-\mathrm{i} d i k_n) \tilde{b}_{k_n}^* a_i))\exp(\sum_{n, n^\prime=1}^N \tilde{b}_{k_n^\prime}^* (2J\cos(d k_n) - \omega ) \delta_{k_n, k_{n^\prime}} \tilde{b}_{k_n})\\
		&\sim \int \mathcal{D}\{a, a^*\} \mathrm{e}^{-S_a} \exp(g^2 \sum_{n=1}^N \sum_{i, j=1}^N \frac{ a_i^* a_j \exp(\mathrm{i}d k_n(i - j))}{\omega - 2 J\cos(k_n d)}),
	\end{align}
	where we denoted by \(S_a\) the qubit on-site term of the action. The effective action of the qubit system and the corresponding effective Hamiltonian are therefore
	\begin{align}
	    S_\mathrm{eff}^{q}(\omega) &=  \omega \sum_{i=1}^N a_i^\dagger a_i - \sum_{i, j=1}^N a_i^\dagger a_j\left(\delta_{i, j} \omega_i + g^2 \sum_{n=1}^N  \frac{  \exp(\mathrm{i}d k_n(i - j))}{\omega - 2 J\cos(k_n d)}\right),\\
	    H_\mathrm{eff}^{ q}(\omega) &= \sum_{i, j=1}^N a_i^\dagger a_j\left(\delta_{i, j} \omega_i + g^2 \sum_{n=1}^N \frac{  \exp(\mathrm{i}d k_n(i - j))}{\omega - 2 J\cos(k_n d)}\right). \label{eq:h_eff_q}
	\end{align}
The calculation is straightforwardly extended to a model with an arbitrary number of photon sites between those sites that are coupled to qubits. When this number is large, we can go to the continuum limit and replace the momentum sum in Eq.~\eqref{eq:h_eff_q} by an integral.  Linearizing the dispersion relation and taking the continuum limit, we recover the noninteracting effective qubit wQED Hamiltonian of Refs.~\cite{lalumiere_input-output_2013, caneva_quantum_2015}.

	\section*{Calculating transmission coefficients and localization lengths}
	
	In this section of the Supplementary Material, we present details of transfer-matrix calculations of transmission coefficients and localization lengths. 
	
	Due to the tridiagonal form
	\begin{align}
		H_{i, j} &= H_{i, j} \delta_{i, j} + H_{i, j} \delta_{\langle i, j + 1\rangle}
	\end{align}
	of the effective photonic Hamiltonian, the corresponding Schr\"odinger equation can be written as
	\begin{align}
		H_{i, j}\psi_j &= \omega \psi_i\\
		\Rightarrow \omega \psi_i &= H_{i, i} \psi_i + \left( H_{i, i + 1} \psi_{i + 1} + H_{i, i - 1} \psi_{i - 1}\right)\\
		\Leftrightarrow \psi_{i + 1} &= \frac{1}{H_{i, i + 1}} \left[ \left(\omega - H_{i, i} \right) \psi_i - H_{i, i - 1} \psi_{i - 1}\right].
	\end{align}
	This allows us to determine wave function coefficients iteratively with the matrix equation
	\begin{align}
	\begin{pmatrix}
		\psi_{i + 1}\\
		\psi_{i}
	\end{pmatrix} =
	\begin{pmatrix}
		\frac{\omega - H_{i, i}}{H_{i, i + 1}} & -\frac{H_{i, i - 1}}{H_{i, i + 1}}\\
		1  & 0
	\end{pmatrix}
	\begin{pmatrix}
		\psi_i\\
		\psi_{i - 1}
	\end{pmatrix},
	\end{align}
	given an initial condition on sites \(\psi_i\), \(\psi_{i - 1}\). From the wave function components on each site, we can calculate an effective localization length \(\xi_i\) that describes the exponential decay of a wave function,
	\begin{align}
		|\psi_i| &\sim \exp(\pm \frac{i \cdot d}{2\xi_i})
	\end{align}
	with the lattice spacing \(d\) (we set \(d:= 1\) in the following). It follows
	\begin{align*}
		\xi &= \lim_{i \rightarrow \infty} \frac{i}{2\log(|\psi_i|)},
	\end{align*}
	where the limit should be chosen such that the localization length converges. 
	
	From the localization length we can find the power transmission coefficient by
	\begin{align}
		T_N := \left| \frac{\psi_{N+1}}{\psi_{0}}\right|^2 \,,
	\end{align}
	where we introduced the sites \(\psi_{N+1}\) and \(\psi_0\).
 We can therefore determine \(T_N\) from \(\xi\) according to
	\begin{align}
		T_N &= \left| \frac{\psi_{N+1}}{\psi_0}\right|^2 = \left| \frac{\exp(-\frac{N}{2\xi_N})}{1}\right|^2,\\
		\Rightarrow T_N &= \exp(-\frac{N}{\xi_N}) \qquad \Leftrightarrow \qquad \frac{1}{\xi_N} = -\frac{\log(T_N)}{N}.
	\end{align}
	
	Alternatively, we can directly calculate transmission coefficients and obtain effective localization lengths from
	\begin{align}
	    \frac{1}{\xi_N} = - \frac{1}{N} \left\langle \log(T_N) \right\rangle.   
	\end{align}
	In order to define the transmission coefficient for a given incident energy, we couple the system of sites \(i \in [1, N]\) to leads extending from \((-\infty, 0]\) and \([N+1, \infty)\) 
	(see for example Ref.~\cite{paper:transmission}). This yields the Hamiltonian
	\begin{align}
		H_\mathrm{coupled} = &\sum_{j=1}^N \omega_j a_j^\dagger a_j + J\sum_{i, j}^N \delta_{\langle i, j\rangle} b^\dagger_i b_{j}+ g \sum_{i=1}^N \left( b_{i}^\dagger a_i + a_i^\dagger b_i \right) \nonumber \\
		&+ J_\mathrm{lead} \sum_{i, j\in \mathbb{N}\setminus [1, N]} d^\dagger_i \delta_{\langle i, j \rangle} d_j + c_\mathrm{L}\left( d^\dagger_0 b_1 + b_1^\dagger d_0 \right)+ c_\mathrm{R} \left(d_{N+1}^\dagger b_N + b_N^\dagger d_{N+1}\right),
		\label{eq:H-coupled}
	\end{align}
	where \(d_j\), \(d_j^\dagger\) are the annihilation and creation operators in the leads, while \(c_\mathrm{L}\) and \(c_\mathrm{R}\) are the couplings between the system and the left / right lead. For our simulations, \(c_{\rm L}  = c_{\rm R} = J_{\rm lead}  = J\). 
	For simplicity, we assume in Eq.~(\ref{eq:H-coupled}) and below that the spacing $d_\gamma$ between the photon sites is the same as the spacing $d$ between the qubits. The analysis is straightforwardly generalized to the case of arbitrary integer $d/d_\gamma$. (The results presented in the main text of the paper are obtained for $d/d_\gamma = 2$.)
	
	Integrating out the qubits, we find the effective photonic Hamiltonian involving the leads:
	\begin{align}
		H_\text{eff, coupled} = &\sum_{i, j=1}^N b_i^\dagger \left(\delta_{i, j} \frac{g^2}{\omega_i - \omega} + J \delta_{\langle i, j \rangle} \right)b_j \\
		&+ J_\mathrm{lead} \sum_{i, j\in \mathbb{N}\setminus [1, N]} d^\dagger_i \delta_{\langle i, j \rangle} d_j\\
		 &+ c_\mathrm{L}\left( d^\dagger_0 b_1 + b_1^\dagger d_0 \right)+ c_\mathrm{R} \left(d_{N+1}^\dagger b_N + b_N^\dagger d_{N+1}\right).
	\end{align}
	The transfer-matrix can be now evaluated in a standard way:
	\begin{align}
			\begin{pmatrix}
			\psi_{N + 2}\\
			\psi_{N + 1}
		\end{pmatrix} &= 
		\begin{pmatrix}
			\frac{\omega }{J_\mathrm{lead}} & - \frac{c_\mathrm{R}}{ J_\mathrm{lead}}\\
			1 & 0
		\end{pmatrix}		
		\begin{pmatrix}
			\frac{\omega -\varepsilon_N}{c_\mathrm{R}} & - \frac{J}{c_\mathrm{R}}\\
			1 & 0
		\end{pmatrix}
		\begin{pmatrix}
			\psi_{N}\\
			\psi_{N - 1}
		\end{pmatrix}\\
		&= T_\mathrm{right} \underbrace{ \left(\prod_{i=2}^{N - 1} Q_i \right)}_{=:T_\mathrm{sys}} 
		\begin{pmatrix}
			\frac{(\omega - \varepsilon_1)}{J} & - \frac{c_\mathrm{L}}{J}\\
			1 & 0
		\end{pmatrix} \begin{pmatrix}
			\frac{\omega}{c_\mathrm{L}} & - \frac{J_\mathrm{lead}}{c_\mathrm{L}}\\
			1 & 0
		\end{pmatrix}\begin{pmatrix}
			\psi_0\\
			\psi_{-1}
		\end{pmatrix}\\
		&= T_\mathrm{right} T_\mathrm{sys} T_\mathrm{left}\begin{pmatrix}
			\psi_0\\
			\psi_{-1}
		\end{pmatrix},\\
		Q_i &:= \begin{pmatrix}
			\frac{\omega - \varepsilon_i}{J} & -1\\
			1 & 0
		\end{pmatrix},\\
		\varepsilon_i &:= \frac{g^2}{\omega_i - \omega} \,.
	\end{align}
	As an initial condition, we choose a plane wave of specified momentum \(k = \arccos(\omega / 2J) / d\) in the leads (corresponding to the frequency \(\omega\)) and convert it to coefficients:
	\begin{align}
		\psi_n &= A_n \exp(\mathrm{i}kdn) + B_n \exp(-\mathrm{i}kdn) \,, \\
		\begin{pmatrix}
			\psi_{n + 1}\\
			\psi_n
		\end{pmatrix} &=\underbrace{
		\begin{pmatrix}
			\exp(\mathrm{i}kd(n + 1)) & \exp(-\mathrm{i}kd(n + 1))\\
			\exp(\mathrm{i}kdn) & \exp(-\mathrm{i}kdn)
		\end{pmatrix}}_{M_{n}(k)}
		\begin{pmatrix}
			A_n\\
			B_n
		\end{pmatrix}.
	\end{align}
	If we consider an incident wave coming from the left lead, we know that there is only a right-moving wave to the right of the system. We therefore set
	\begin{align}
		\begin{pmatrix}
			\psi_{N+2}\\
			\psi_{N +1}
		\end{pmatrix}(k) = M_{N + 1}(k) \begin{pmatrix}
			A_{N + 1}\\
			B_{N + 1}
		\end{pmatrix}= M_{N + 1}(k)
		\begin{pmatrix}
			1\\
			0
		\end{pmatrix}
	\end{align}
	and write with \(T = T_\mathrm{right} T_\mathrm{sys} T_\mathrm{left}\)
	\begin{align}
		\begin{pmatrix}
			A_{-1}\\
			B_{-1}
		\end{pmatrix}(k)&=
		M_{-1}^{-1}(k) \begin{pmatrix}
			\psi_{0}\\
			\psi_{-1}
		\end{pmatrix}(k)\\
		&=
		M_{-2}^{-1}(k) T^{-1}(k)\begin{pmatrix}
			\psi_{N+2}\\
			\psi_{N+1}
		\end{pmatrix}(k)\\
		&=M_{-2}^{-1}(k) T^{-1}(k) M_{N + 1}(k)
		\begin{pmatrix}
			1\\
			0
		\end{pmatrix}.
	\end{align}
	Transmission and reflection coefficients \(t\) and \(r\) are defined as
	\begin{align}
		t(k)&= \left|\frac{1}{A_{-1}(k)}\right|^2 \,,
		\\ 
		r(k)&= \frac{\left|B_{-1}(k)\right|^{2}}{\left|A_{-1}(k)\right|^2}.
	\end{align}
We use the first of these equations to determine \(t(k)\).

\end{document}